\documentclass[draftclsnofoot,onecolumn]{IEEEtran}
\ifCLASSINFOpdf
\else
\fi
\usepackage{amsbsy,amscd,amsfonts,amsgen,amsmath,amsopn,amssymb,amstext,amsthm,amsxtra}
\usepackage{graphicx}\usepackage{subfigure}
\usepackage{verbatim}
\usepackage{citesort}
\usepackage{url}
\usepackage{color}
\usepackage{epstopdf}
\usepackage{caption}
\usepackage{changes}
\usepackage{bbm}
\graphicspath{{./} {img/}}

\IEEEoverridecommandlockouts

\newcounter{cnt01}
\newcounter{cnt02}
\addtocounter{cnt01}{2}
\addtocounter{cnt02}{3}

\begin{document}
\title{Compressive Imaging via Approximate Message Passing with Image Denoising}
\author{Jin~Tan,~\IEEEmembership{Student Member,~IEEE,}
Yanting~Ma,~\IEEEmembership{Student Member,~IEEE,}\\
and Dror~Baron,~\IEEEmembership{Senior Member,~IEEE}
\thanks{This work was supported in part by the National Science Foundation under Grant CCF-1217749 and in part by the U.S. Army Research Office under Grants W911NF-04-D-0003 and W911NF-14-1-0314.}
\thanks{Jin Tan, Yanting Ma, and Dror Baron are with the Department of Electrical and Computer Engineering, NC State University, Raleigh, NC 27695. E-mail: \{jtan, yma7, barondror\}@ncsu.edu.}
}

%
\maketitle
\thispagestyle{empty}
\newcommand{\xhat}{\widehat{\mathbf{x}}}
\newcommand{\xhati}{\widehat{x}_i}
\def\x{{\mathbf x}}
\def\L{{\cal L}}
\begin{abstract}
We consider compressive imaging problems, where images are reconstructed 
from a reduced number of linear measurements.
Our objective is to improve 
over existing compressive imaging algorithms in terms of both reconstruction
error and runtime. To pursue our objective, we propose compressive imaging
algorithms that employ the approximate message passing (AMP) framework. 
AMP is an iterative signal reconstruction algorithm that performs scalar
denoising at each iteration; in order for AMP to reconstruct the original
input signal well, a good denoiser must be used.
We apply two wavelet based image denoisers within AMP.
The first denoiser is the ``amplitude-scale-invariant Bayes estimator" (ABE),
and the second is an adaptive Wiener filter; we call our 
AMP based algorithms for compressive imaging AMP-ABE and AMP-Wiener.
Numerical results show that both AMP-ABE and AMP-Wiener 
significantly improve over the state of the art in terms of runtime.
In terms of reconstruction quality, AMP-Wiener offers lower mean square
error (MSE) than existing compressive imaging algorithms.
In contrast, AMP-ABE has higher MSE, because ABE does not denoise as
well as the adaptive Wiener filter.
\end{abstract}
\IEEEpeerreviewmaketitle

\section{Introduction}
\label{sec:intro}
\subsection{Motivation}
\label{subsec:motivation}

Compressed sensing (CS)~\cite{CandesRUP,DonohoCS} has
sparked a tremendous amount of research activity in recent years, 
because it performs signal acquisition and processing
using far fewer samples than required by the Nyquist rate.
Breakthroughs in CS have the potential to 
greatly reduce the sampling
rates in numerous signal processing applications such as 
cameras~\cite{takhar2006},
medical scanners, 
fast analog to digital converters~\cite{Tropp2006random}, 
and high speed radar~\cite{BaraniukCS2007}. 

Compressed sensing has been used in compressive imaging, where
the input signal is an image, and the goal is to acquire the
image using as few measurements as possible.
Acquiring images in a compressive manner requires
less sampling time than conventional imaging technologies.
Applications of compressive imaging appear in 
medical imaging~\cite{Blahut2004theory,Natterer2001,Liang2000}, 
seismic imaging~\cite{Claerbout1985imaging}, 
and hyperspectral imaging~\cite{Rajwade2013,Willett2014}.

\subsection{Related work}
\label{subsec:relatedWork}

Many compressive imaging algorithms have been proposed in the literature. For example, Som and Schniter~\cite{Som2012} modeled the structure of the wavelet coefficients by a hidden Markov tree (HMT), and applied a turbo scheme that alternates between inference on the HMT structure with standard belief propagation and inference on the compressed sensing  measurement structure with the generalized approximate message passing algorithm. 
He and Carin~\cite{BCS2008spike} proposed a hierarchical
Bayesian approach with Markov chain Monte Carlo (MCMC) for natural image reconstruction.
Soni and Haupt~\cite{Soni2012} exploited a hierarchical dictionary learning method~~\cite{Jenatton2010} and assumed that projecting images onto the learned dictionary will yield tree-sparsity, and therefore the nonzero supports of the dictionary can be identified and estimated accurately by setting an appropriate threshold.

However, existing compressive imaging algorithms may either not achieve good reconstruction quality or not be fast enough. Therefore, in this paper, we focus on a variation of a fast and effective algorithm called approximate message passing (AMP)~\cite{DMM2009} to improve over the prior art.
AMP is an iterative signal reconstruction algorithm that performs scalar denoising within each iteration,
and proper selection of the denoising function used within AMP is needed to obtain better reconstruction quality.
One challenge in applying image denoisers within AMP is that it may be hard to compute the so-called ``Onsager reaction term"~\cite{Thouless1977,DMM2009} in the AMP iteration steps. The Onsager reaction term includes the derivative of the image denoising function, and thus if an image function does not have a convenient closed form, then the Onsager reaction term may be difficult to
compute.

Dictionary learning is an effective technique that has attracted a great deal of attention in image denoising. Dictionary learning based methods~\cite{Zhoul2011,Fang2013}
generally achieve lower reconstruction error than wavelet-based methods.
However, the learning procedure requires a large amount of training images,
and may involve manual tuning. Owing to these limitations, our main
focus in this paper is to integrate relatively simple and fast image denoisers
into compressive imaging reconstruction algorithms. 

Dabov et al.~\cite{Dabov2007} developed an image denoising strategy that employs collaborative filtering in a sparse 3-D transform domain, and they offered an efficient implementation that achieves favorable denoising quality.
Other efficient denoising schemes include wavelet-based methods.
A typical wavelet-based image denoiser proceeds as follows:
({\em i}) apply a wavelet transform to the image and obtain wavelet coefficients; 
({\em ii}) denoise the wavelet coefficients; and
({\em iii}) apply an inverse wavelet transform to the denoised 
wavelet coefficients, yielding a denoised image. 
Two popular examples of denoisers that can be applied to the
wavelet coefficients are
hard thresholding and soft thresholding~\cite{Donoho1994}.
Variations on the thresholding scheme can be found in \cite{Figueiredo2001,Chang2000}; other wavelet-based methods were proposed by Simoncelli and Adelson~\cite{Simoncelli1996}, M{\i}h{\c{c}}ak et al.~\cite{Kivanc1999}, and Moulin and Liu~\cite{Moulin1999}.

\subsection{Contributions}
\label{sec:contrib}

The objective of this paper is to develop compressive imaging algorithms that are fast and reconstruct well.
We apply the ``amplitude-scale-invariant Bayes estimator" (ABE)~\cite{Figueiredo2001} and the adaptive Wiener filter~\cite{Kivanc1999} as image denoisers within AMP.
Our numerical results with these denoisers are promising, showing that our AMP based algorithms run at least 3.5 times faster than the prior art algorithms. Moreover, with a proper choice of image denoiser within AMP, the proposed algorithm also outperforms the prior art algorithms in reconstruction quality. Although we only test for two wavelet-based image denoisers within AMP, we believe that other image denoisers would also work within the AMP framework, and these denoisers need not be wavelet-based.

The remainder of the paper is arranged as follows. We review AMP in Section \ref{sec:review}, 
and describe the image denoisers~\cite{Figueiredo2001,Kivanc1999} that we use within AMP in Section~\ref{sec:Denoisers}.
Numerical results are presented in Section \ref{sec:NumSim}, and the paper concludes with a discussion in Section~\ref{sec:disc}.

\section{Review of Approximate Message Passing}
\label{sec:review}

\subsection{Problem setting}
Before reviewing AMP, let us first model how measurements are obtained in a compressive
imaging system. 

{\bf Matrix channels: }We rearrange the input image $\bf{x}$, which is 
comprised of $N$ pixels, as a column vector of length $N$
(for example, for a $512\times512$ image, $N=512^2$).
Then, we multiply $\bf{x}$ by a known measurement 
matrix~${\bf A}\in\mathbb{R}^{M\times N}$, which
has $M$ rows (typically~$M<N$).
Finally, the measurements are corrupted by independent and identically distributed (i.i.d.) zero-mean Gaussian noise~${\bf z}$, 
\begin{eqnarray}
\label{eq:basicSystem}
\mathbf{y =A x +z}.
\end{eqnarray}
We observe the noisy vector $\mathbf{y}\in\mathbb{R}^M$, 
and want to estimate and reconstruct the original input signal $\mathbf{x}$ from $\mathbf{y}$ and $\mathbf{A}$. 

{\bf Scalar channels: }
We now define scalar channels, and will describe in Section~\ref{subsec.AMP} how the matrix channel is converted to scalar channels in AMP. 
In scalar channels, the noisy observations obey
\begin{equation}
\label{eq.scalar}
q_i=x_i+v_i,
\end{equation}
$i\in\{1,2,\ldots,N\}$,
the subscript~$(\cdot)_i$ denotes the $i$-th component of a vector (denoted by a lower case letter with bold font), and~${\bf x},{\bf q}\in\mathbb{R}^N$ are the input signal and the noisy observations, respectively. The noise~${\bf v}$ is i.i.d. Gaussian, $v_i\sim\mathcal{N}(0,\sigma^2)$.
Note that we use different notations for the noise and observations in matrix channels and scalar channels. The main difference between the two types of channels is that 
the observations~${\bf y}$ in the matrix channel contain linear combinations of the entries of ${\bf x}$.

\subsection{Algorithmic framework}
\label{subsec.AMP}

We are now ready to describe AMP~\cite{DMM2009}, which is an iterative signal reconstruction algorithm in matrix channels.
Consider a matrix channel model~\eqref{eq:basicSystem} where the signal distribution follows~${x}_i\sim f_{X}$ and the noise is i.i.d. Gaussian. 
The entries of the measurement matrix~${\bf A}$ are i.i.d.~$\mathcal{N}(0,\frac{1}{M})$ distributed, and thus the columns of the matrix have unit $\ell_2$-norm, on average.
AMP~\cite{DMM2009} proceeds iteratively according to
\begin{align}
{\bf x}^{t+1}&=\eta_t({\bf A}^T{\bf r}^t+{\bf x}^t)\label{eq.AMPiter1},\\
{\bf r}^t&={\bf y}-{\bf Ax}^t+\frac{1}{R}{\bf r}^{t-1}
\langle\eta_{t-1}'({\bf A}^T{\bf r}^{t-1}+{\bf x}^{t-1})\rangle\label{eq.AMPiter2},
\end{align}
where~${\bf A}^T$ is the transpose of ${\bf A}$, $R=M/N$ represents the measurement rate, $\eta_t(\cdot)$ is a denoising function at the $t$-th iteration, $\eta_t'({\bf s})=\frac{\partial}{\partial {\bf s}}\eta_t({\bf s})$, and~$\langle{\bf u}\rangle=\frac{1}{N}\sum_{i=1}^N u_i$
for some vector~${\bf u}=(u_1,u_2,\ldots,u_N)$.
The last term in equation~\eqref{eq.AMPiter2} is called the ``Onsager reaction term"~\cite{Thouless1977,DMM2009} in statistical physics.
In the~$t$-th iteration, we obtain the vectors~${\bf x}^t\in\mathbb{R}^N$ and~${\bf r}^t\in\mathbb{R}^M$. 
We highlight that the vector~${\bf A}^T{\bf r}^t+{\bf x}^t\in\mathbb{R}^N$ in~\eqref{eq.AMPiter1} can be regarded as noisy measurements of~${\bf x}$ in the~$t$-th iteration with noise variance~$\sigma_t^2$, and therefore the denoising function~$\eta_t(\cdot)$ is performed on a scalar channel~\eqref{eq.scalar}. Let us denote the equivalent scalar channel at iteration~$t$ by
\begin{equation}
{\bf q}^t = {\bf A}^T{\bf r}^t +{\bf x}^t= {\bf x} + {\bf v}^t,
\label{eq:scalar_t}
\end{equation}
where~$v^t_i\sim\mathcal{N}(0,\sigma_t^2)$.
The asymptotic performance of AMP can be characterized by a {state evolution} (SE) formalism:
\begin{equation}
\sigma^2_{t+1}=\sigma^2_z+\frac{1}{R}E\left[\left( \eta_t\left( X+\sigma_{t}W \right)-X \right)^2\right]
\label{eq.SE},
\end{equation}
where the random variables~$W\sim\mathcal{N}(0,1)$ and~$X\sim f_{X}$.
Formal statements about SE appear in Bayati and Montanari~\cite{Bayati2011}.
Note that SE~\eqref{eq.SE} tracks the noise variance for AMP iterations, but the noise variance~$\sigma_{t+1}^2$ need not necessarily be
the smallest possible, unless in each iteration 
the denoiser~$\eta_t(\cdot)$ achieves the minimum mean square error (MMSE). 
We note in passing that we have proposed a denoiser called ``MixD"~\cite{MTKB2014ITA} that achieves the MMSE of scalar channels~\eqref{eq.scalar}, and thus applying MixD within AMP achieves the MMSE of matrix channels~\eqref{eq:basicSystem}.
On the other hand, it is unrealistic to expect existing image denoisers to achieve the MMSE, because the statistical distribution of natural images has yet to be determined. That said, running AMP with good  image denoisers that achieve lower mean square error (MSE) may yield lower MSE in compressive imaging problems.

Finally, SE theoretically characterizes the noise variance $\sigma_t^2$ of the scalar channel at each iteration. However, the MSE performance of image denoisers cannot be characterized theoretically. Therefore, we must estimate the effective Gaussian noise level $\sigma^2_t$ empirically in each AMP iteration. The estimated noise variance $\widehat{\sigma}^2_t$ can be calculated as~\cite{Montanari2012}:
\begin{equation}
\widehat{\sigma}^2_t=\frac{1}{M}\sum_{i=1}^M (r^t_i)^2,\label{eq.sigma_t}
\end{equation}
where ${\bf r}^t$ is defined in (\ref{eq.AMPiter2}).

\section{Image Denoising within AMP}
\label{sec:Denoisers}

In this section, 
we describe how wavelet-based image denoisers are applied within AMP, and then
outline two image denoisers that were proposed by Figueiredo and 
Nowak~\cite{Figueiredo2001} and M{\i}h{\c{c}}ak et al.~\cite{Kivanc1999},
respectively.

\subsection{Wavelet transforms in AMP}
\label{subsec.waveAMP}

In image processing, one often computes the wavelet coefficients~\cite{Mallat1999book} of images, applies some signal 
processing technique to the wavelet coefficients, and finally applies the inverse wavelet transform to the processed coefficients to obtain processed images. 
We now show how image denoising can be performed within AMP in the 
wavelet domain.
Let us denote the wavelet transform by~$\mathcal{W}$ and the inverse wavelet transform by~$\mathcal{W}^{-1}$. By applying the wavelet transform 
to a vectorized image signal~${\bf x}$ (a 2-dimensional wavelet transform
is used),
we obtain the wavelet coefficient vector~${\bf\theta_x}=\mathcal{W}{\bf x}$. Conversely,~${\bf x}=\mathcal{W}^{-1}{\bf\theta_x}$. Therefore, the matrix 
channel~\eqref{eq:basicSystem} becomes
${\bf y=A}\mathcal{W}^{-1}{\bf \theta_x+z}$,
where~${\bf A}\mathcal{W}^{-1}$ can be regarded as a new matrix in the matrix channel~\eqref{eq:basicSystem} and~${\bf\theta_x}$ as the corresponding input signal. 

Let us express the AMP iterations~(\ref{eq.AMPiter1},~\ref{eq.AMPiter2}) for settings where the matrix is~${\bf A}\mathcal{W}^{-1}$,
\begin{eqnarray}
\theta^{t+1}_{\bf x}&=&\eta_t(({{\bf A}\mathcal{W}^{-1}})^T{\bf r}^t+\theta^t_{\bf x})\label{eq.waveAMP1},\\
{\bf r}^t&=&
{\bf y}-({\bf A}\mathcal{W}^{-1})\theta^t_{\bf x}\nonumber\\
&&+\frac{1}{R}{\bf r}^{t-1}
\langle\eta_{t-1}'(({{\bf A}\mathcal{W}^{-1}})^T{\bf r}^{t-1}+\theta^{t-1}
_{\bf x})\rangle\nonumber\\
&=&
{\bf y}-{\bf Ax}^t\nonumber\\
&&+\frac{1}{R}{\bf r}^{t-1}
\langle\eta_{t-1}'(({{\bf A}\mathcal{W}^{-1}})^T{\bf r}^{t-1}+\theta^{t-1}_{\bf x})\rangle\label{eq.waveAMP2}.
\end{eqnarray}
Because the wavelet transform matrix is orthonormal, i.e.,~$\mathcal{W}\mathcal{W}^T={\bf I}=\mathcal{W}\mathcal{W}^{-1}$, it can be shown 
that~$({\bf A}\mathcal{W}^{-1})^T=\mathcal{W}{\bf A}^T$. 
Therefore, the input of the denoiser~$\eta_t(\cdot)$~\eqref{eq.waveAMP1} becomes
\begin{equation}
({{\bf A}\mathcal{W}^{-1}})^T{\bf r}^t+\theta^t_{\bf x}=
\mathcal{W}{\bf A}^T{\bf r}^t + \theta^t_{\bf x}
=\mathcal{W}{\bf A}^T{\bf r}^t + \mathcal{W}{\bf x}^t
=\mathcal{W}{\bf q}^t,
\end{equation}
where~${\bf q}^t$~\eqref{eq:scalar_t} is the noisy image at iteration~$t$, 
and~$\mathcal{W}{\bf q}^t$ is the wavelet transform 
applied to the noisy image. 

With the above analysis of the modified AMP~(\ref{eq.waveAMP1},~\ref{eq.waveAMP2}), we formulate a compressive imaging procedure as follows. Let us denote the the wavelet transform of the scalar channel~\eqref{eq:scalar_t} by
\begin{equation}
\theta_{\bf q}^t = \theta_{\bf x}+\theta_{\bf v}^t,
\label{eq:scalar_theta}
\end{equation}
where~$\theta_{\bf q}^t = \mathcal{W}{\bf q}^t$, $\theta_{\bf x}=\mathcal{W}{\bf x}$, and $\theta_{\bf v}^t=\mathcal{W}{\bf v}^t$.
First,~${\bf r}^t$ and ${\bf x}^t$ are initialized to all-zero vectors.
Then, at iteration~$t$ the algorithm proceeds as follows,
\begin{enumerate}
\item Calculate the residual term~${\bf r}^t$.
\item Calculate the noisy image~${\bf q}^t={\bf A}^T{\bf r}^t+{\bf x}^t$, and apply the wavelet transform~$\mathcal{W}$ to the noisy image~${\bf q}^t$ to obtain wavelet coefficients~$\theta_{\bf q}^t$, which are the inputs of the scalar denoiser~$\eta_t(\cdot)$ in~\eqref{eq.waveAMP1}.
\item Apply the denoiser~$\eta_t(\cdot)$ to the wavelet coefficients~$\theta_{\bf q}^t$, and obtain denoised coefficients~$\theta^{t+1}_{\bf x}$.
\item Apply the inverse wavelet transform~$\mathcal{W}^{-1}$ to the coefficients~$\theta_{\bf x}^{t+1}$ to obtain the estimated image~${\bf x}^{t+1}$, which is used to compute the residual term in the next iteration.
\end{enumerate}

\subsection{Image denoisers}
\label{subsec.FN_denoiser}


We choose to denoise the wavelet coefficients
using scalar denoisers proposed by 
Figueiredo and Nowak~\cite{Figueiredo2001} and M{\i}h{\c{c}}ak et al.~\cite{Kivanc1999}, respectively, because these two denoisers are simple to implement while revealing promising numerical results (see Section~\ref{sec:NumSim}). 
We call the algorithm where ABE~\cite{Figueiredo2001} is utilized within AMP ``AMP-ABE," and the algorithm where the adaptive Wiener filter~\cite{Kivanc1999} is utilized ``AMP-Wiener."
In both algorithms, the variance of the noise~$\sigma_t^2$ in the noisy image is obtained using~\eqref{eq.sigma_t}. Because we use an orthonormal wavelet transform, the noise variance in the wavelet domain is equal to that in the image domain.
Although we only show how to employ two image denoisers within AMP, they serve as a proof of concept that other  image denoisers could also be applied within AMP, possibly leading to further improvements in both image reconstruction quality and runtime.

\subsubsection{Amplitude-scale-invariant Bayes estimator}
\label{subssub:ABE}

\begin{figure}[t]
\vspace*{-1mm}
\begin{center}
\includegraphics[width=80mm]{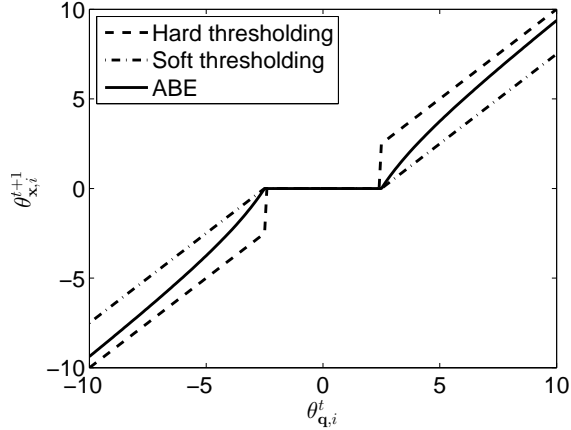}
\end{center}
\vspace*{-6mm}
\caption{\small\sl 
Comparison of ABE with hard and soft thresholding.}
\label{fig.denoisers}
\end{figure}

Figueiredo and Nowak's denoiser~\cite{Figueiredo2001} is an amplitude-scale-invariant Bayes estimator (ABE), and it is a scalar function. More specifically, for each noisy wavelet coefficient~$\theta_{{\bf q},i}^t$~\eqref{eq:scalar_theta}, the estimate of~$\theta_{{\bf x},i}$ for the next iteration is
\begin{equation}
{\theta}_{{\bf x},i}^{t+1}=\eta_t(\theta_{{\bf q},i}^t)=\frac{\left((\theta_{{\bf q},i}^t)^2-3\sigma_t^2\right)_+}{\theta_{{\bf q},i}^t},
\label{eq:eta_theta}
\end{equation}
where~$\sigma_t^2$ is the noise variance of the scalar channel~\eqref{eq:scalar_t} at the $t$-th AMP iteration,
and $(\cdot)_+$ is a function such that $(u)_+=u$ if $u>0$ and 
$(u)_+=0$ if $u\le0$.
Note that because the wavelet transform matrix~$\mathcal{W}$ is orthonormal, the variance of the noise ${\bf v}^t$~\eqref{eq:scalar_t} is equal to the variance of~$\theta_{\bf v}^t$~\eqref{eq:scalar_theta}.

Figure~\ref{fig.denoisers} illustrates the ABE function. It can be 
seen that ABE offers a compromise between the hard thresholding,
i.e., $\eta(u,\tau) = u\cdot \mathbbm{1}_{(u>|\tau|)}$,
where $\mathbbm{1}_{\{\cdot\}}$ 
denotes an indicator function,
and soft thresholding, i.e., $\eta(u,\tau) = \text{sign}(u)\cdot(|u|-\tau)_+$, proposed by
Donoho and Johnstone~\cite{Donoho1994}. 
ABE is convenient to utilize, because there is no need to tune the thresholding values,\footnote{We note in passing that Mousavi et al.~\cite{Mousavi2013} proposed to tune the soft threshold automatically.} and ABE has been shown to outperform both hard and soft thresholding methods for image denoising~\cite{Figueiredo2001}.

The ABE function is continuous and differentiable except for two points ($\theta_{{\bf q},i}^t=\pm\sqrt{3}\sigma_t$), and  we calculate the derivative of this denoising function numerically to obtain the Onsager reaction term in~\eqref{eq.AMPiter2}.

\subsubsection{Adaptive Wiener filter}

M{\i}h{\c{c}}ak et al.~\cite{Kivanc1999} proposed a method to estimate the variances of the wavelet coefficients, and then apply the corresponding Wiener filter to each wavelet coefficient. 
The variance of the noisy wavelet coefficient $\theta_{{\bf q},i}^t$ is estimated from its neighboring coefficients. More specifically, a set of $3\times 3$ or $5\times 5$ neighboring coefficients $\mathcal{N}_i$ that is centered at $\theta_{{\bf q},i}^t$ is considered, and the variance of $\theta_{{\bf q},i}^t$ is estimated by averaging the sum of $(\theta_{{\bf q},k}^t)^2$ where $k\in\mathcal{N}_i$. This method of averaging the neighboring coefficients can be regarded as first convolving a $3\times3$ or $5\times5$ mask of all $1$'s with the matrix of squared wavelet coefficients~$\theta_{\bf q}^t$, and then dividing by the normalizing constant $9$ (for a $3\times 3$ mask) or $25$ (for a $5\times 5$ mask).
Other masks can be applied to produce different and possibly better denoising results. For example, we have found that the mask
\begin{equation}
\begin{matrix}
&1&1&1&\\
1&1&2&1&1\\
1&2&3&2&1\\
1&1&2&1&1\\
&1&1&1&
\end{matrix}
\nonumber
\end{equation}
obtains lower MSE than other $5\times5$ masks we have considered.
Recall the scalar channel defined in~\eqref{eq:scalar_theta} where the noise variance is~${\sigma}_t^2$; we estimate the variance of a noisy wavelet coefficient~$\theta_{{\bf q},i}^t$ by~$\widehat{\sigma}_i^2$, and the variance of the true wavelet coefficient~$\theta_{{\bf x},i}^t$ by $\widehat{\sigma}_i^2-\sigma_t^2$.\footnote{We use $\max\{\widehat{\sigma}_i^2-\sigma_t^2,0\}$ to restrict the variance to be non-negative.} 
Therefore, the scaling factor in the Wiener filter~\cite{Wiener1949} is given by $\frac{\widehat{\sigma}_i^2-\sigma_t^2}{(\widehat{\sigma}_i^2-\sigma_t^2) + \sigma_t^2}$, and the adaptive Wiener filter being used as the denoising function can be expressed as follows,
\begin{equation}
{\theta}_{{\bf x},i}^{t+1}=\eta_t(\theta_{{\bf q},i}^t)
=\frac{\widehat{\sigma}_i^2-\sigma_t^2}{(\widehat{\sigma}_i^2-\sigma_t^2) + \sigma_t^2}\theta_{{\bf q},i}^t
=\frac{\widehat{\sigma}_i^2-\sigma_t^2}{\widehat{\sigma}_i^2}\theta_{{\bf q},i}^t.
\label{eq.localWiener}
\end{equation}
Finally, the derivative of this denoising function with respect to $\theta_{{\bf q},i}^t$ is simply the scaling factor $\frac{\widehat{\sigma}_i^2-\sigma_t^2}{\widehat{\sigma}_i^2}$ of the Wiener filter, and so the Onsager reaction term in~\eqref{eq.AMPiter2} can be obtained efficiently.

In standard AMP~\cite{DMM2009}, the denoising function~$\eta_t(\cdot)$ is
separable, meaning that ${\theta}_{{\bf x},i}^{t+1}$ only depends on its
corresponding noisy wavelet coefficient~${\theta}_{{\bf q},i}^t$. In the
adaptive Wiener filter, however, the estimated variance~$\widehat{\sigma}_i^2$
of each noisy wavelet coefficient depends on the neighboring coefficients of
${\theta}_{{\bf q},i}^t$, and so the denoising function 
in~\eqref{eq.localWiener} implicitly depends on the neighboring coefficients 
of ${\theta}_{{\bf q},i}^t$. Therefore, the adaptive Wiener filter 
in~\eqref{eq.localWiener} is not a strictly separable denoising function,
and AMP-Wiener encounters convergence issues. Fortunately, a technique called ``damping"~\cite{Rangan2014ISIT} solves for the convergence problem of AMP-Wiener. Specifically, damping is an extra step in the AMP iteration~\eqref{eq.AMPiter1}; instead of updating the value of~${\bf x}^{t+1}$ by the output of the denoiser~$\eta_t({\bf A}^T{\bf r}^{t}+{\bf x}^{t})$, we assign a weighted sum of~$\eta_t({\bf A}^T{\bf r}^{t}+{\bf x}^{t})$ and~${\bf x}^t$ to~${\bf x}^{t+1}$ as follows,
\begin{equation}
{\bf x}^{t+1} = (1-\lambda)\cdot\eta_t({\bf A}^T{\bf r}^{t}+{\bf x}^{t})+\lambda\cdot{\bf x}^t,
\label{eq.Damping}
\end{equation}
for some constant~$0\le\lambda<1$.
It has been shown by Rangan et al.~\cite{Rangan2014ISIT} that 
sufficient damping ensures the convergence of AMP where the 
measurement matrix~${\bf A}$ is not i.i.d. Gaussian. 
However, we did indeed use i.i.d. Gaussian matrices in our numerical results
in Section~\ref{sec:NumSim}, and damping solved the convergence problem of 
AMP-Wiener, which suggests that damping may be an effective technique when various convergence issues arise in AMP based algorithms.
We note in passing that other techniques such as SwAMP~\cite{Swamp2014} and ADMM-GAMP~\cite{RanganADMMGAMP2015} also solve for the convergence problem in AMP.

\section{Numerical Results}
\label{sec:NumSim}

Having described the AMP algorithm~\cite{DMM2009} and two image denoisers~\cite{Figueiredo2001,Kivanc1999}, in this section we present the numerical results of applying these two denoisers within AMP.


\subsection{Reconstruction quality and runtime}
\label{subsec.runtime}

We compare AMP-ABE and AMP-Wiener with three prior art compressive imaging algorithms, 
({\em i}) Turbo-BG proposed by Som and Schniter~\cite{Som2012};
({\em ii}) Turbo-GM, also by Som and Schniter~\cite{Som2012};
and ({\em iii}) a Markov chain Monte Carlo (MCMC) method by
He and Carin~\cite{BCS2008spike}. 
Both Turbo-BG and Turbo-GM are also message passing based algorithms. However, these two algorithms require more computation than AMP-ABE and AMP-Wiener, because they include two message passing procedures; 
the first procedure solves for dependencies between 
the wavelet coefficients and the second procedure is AMP.
The performance metrics that we use to compare the algorithms 
are runtime and normalized MSE (NMSE),
$\text{NMSE}({\bf x},\widehat{\bf x}) = 10\log_{10} (\|{\bf x - \widehat{x}}\|_2^2/\|{\bf x}\|_2^2)$,
where~$\widehat{\bf x}$ is the estimate of the vectorized input 
image~${\bf x}$.
In all simulations, we use the Haar wavelet transform~\cite{Mallat1999book}.

Let us begin by contrasting the three prior art compressive imaging algorithms based on the numerical results provided in~\cite{Som2012}.
Turbo-BG and Turbo-GM have similar runtimes; 
the NMSE of Turbo-GM is typically 0.5 dB better (lower)
than the NMSE of Turbo-BG.
At the same time, the NMSE of the MCMC algorithm~\cite{BCS2008spike}
is comparable to those of Turbo-BG and Turbo-GM,
but MCMC is 30 times slower than the Turbo 
approaches of Som and Schniter~\cite{Som2012}.
Other algorithms have also been considered for compressive
imaging. For example, compressive sampling
matching pursuit (CoSaMP)~\cite{Cosamp08} requires only half the runtime
of Turbo-GM, but its NMSE is roughly 4 dB worse than that of Turbo-GM;
and model based CS~\cite{BCDH2008} is twice slower than Turbo-GM and
its NMSE is also roughly 4 dB worse.
Therefore, we provide numerical results for Turbo-BG, Turbo-GM, MCMC, and our two proposed AMP based approaches.


{\bf Numerical setting:}\ 
We downloaded 591 images from
``pixel-wise labeled image database v2" at
http://research.\
microsoft.com/en-us/projects/objectclassrecognition, and extracted image patches using the following two methods.
\begin{itemize}
\item Method 1: A $192\times 192$ patch is extracted from the upper left 
corner of each image, and then the patch is resized to $128\times 128$;
this image patch extraction method was used by Som and 
Schniter~\cite{Som2012}.
\item Method 2: A $192\times 192$ patch is extracted from the upper left corner of each image without resizing.
\end{itemize}
The measurement matrix~${\bf A}\in\mathbb{R}^{M\times N}$ is 
generated with i.i.d. Gaussian entries distributed as~$\mathcal{N}(0,\frac{1}{M})$; each column is then normalized to have unit norm. 
For $128\times128$ patches extracted by Method 1, the number of 
measurements~$M=5,000$, which is identical to the numerical setting by Som 
and Schniter~\cite{Som2012}.
For $192\times192$ patches extracted by Method 2, the number of measurements~$M=11,059$, i.e., the measurement rate is $0.3$. 
In both methods, 
the measurements~${\bf y}$ are noiseless, i.e.,~${\bf y=Ax}$. 
Finally, we set the number of AMP iterations to be 30 and the damping
constant~$\lambda$ (\ref{eq.Damping})
for AMP-Wiener to be $0.1$.

\begin{table}[t]
\vspace*{-0mm}
\centering
\begin{tabular}{| c | c | c |}
\hline
Algorithm   & NMSE (dB) & Runtime (sec) \\
\hline 
 Turbo-BG~\cite{Som2012} & -20.37 & 12.39 \\

  Turbo-GM~\cite{Som2012} & -20.72 & 12.47 \\

  MCMC~\cite{BCS2008spike} & -20.31 & 423.15\\

  AMP-ABE & -19.30 & 3.39 \\
    
AMP-Wiener & -21.00 &  3.34 \\
\hline
\end{tabular}     
\caption{\small\sl 
NMSE and runtime averaged over 591 image patches: a $192\times192$ patch from the upper left corner of each image is first extracted, and then resized to $128\times128$.
The number of measurements~$M=5,000$, 
and the measurements are noiseless.}
\label{tb.resize}
\end{table}

\begin{figure}[t]
\vspace*{-2mm}
\begin{center}
\includegraphics[width=80mm]{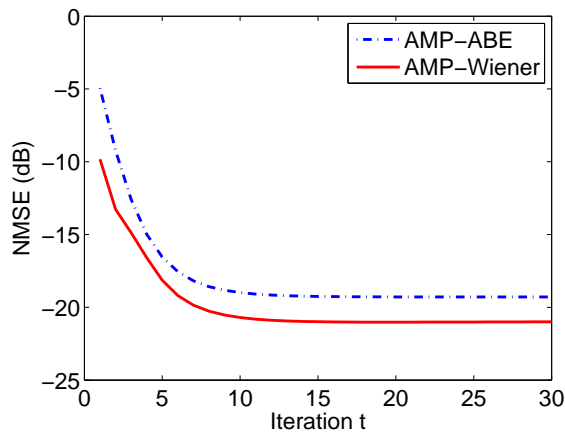}
\end{center}
\vspace*{-0mm}
\caption{\small\sl 
Average NMSE over 591 images from AMP iteration 1 to iteration 30.
Image patches are extracted by Method 1: a $192\times192$ patch from the upper left corner of each image is first extracted, and then resized to $128\times128$.}
\label{fig.MSE_evolve}
\end{figure}

\begin{figure}[t]
\vspace*{-0mm}
\begin{center}
\includegraphics[width=80mm]{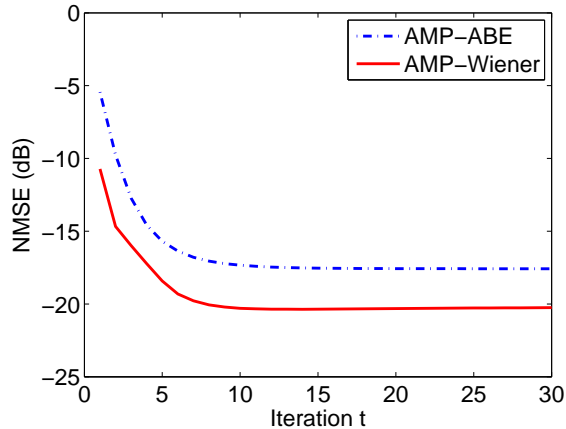}
\end{center}
\vspace*{-0mm}
\caption{\small\sl 
Average NMSE over 591 images from AMP iteration 1 to iteration 30. Image patches are extracted by Method 2: a $192\times 192$ patch is extracted from the upper left corner of each image.}
\label{fig.MSE_evolveb}
\end{figure}

{\bf Result 1:}\
Tables \ref{tb.resize} and~\ref{tb.192} show the NMSE and runtime 
averaged over the 591 image patches that are extracted by Methods~1 
and~2, respectively. Runtime is measured in seconds on a Dell OPTIPLEX~9010
running an Intel(R) $\text{Core}^{\text{TM}}$ i7-860 with 16GB RAM, 
and the environment is Matlab R2013a. Figures~\ref{fig.MSE_evolve} 
and~\ref{fig.MSE_evolveb} complement Tables~\ref{tb.resize} 
and~\ref{tb.192}, respectively, by plotting the average NMSE over 591 images from iteration~1 to iteration~30.

It can be seen from Table~\ref{tb.resize} that the NMSE of AMP-Wiener is the best (lowest) among all the algorithms compared. 
At the same time, AMP-Wiener 
runs approximately 3.5 times faster than the Turbo approaches of
Som and Schniter~\cite{Som2012}, and 120 times faster than
MCMC~\cite{BCS2008spike}. 
Although AMP-ABE does not outperform the competing algorithms in 
terms of NMSE, it runs as fast as AMP-Wiener. 

Table~\ref{tb.resize} presents the runtimes of AMP-ABE and AMP-Wiener for 
image patches extracted by Method 1 with 30 iterations. However, we can 
see from Figure~\ref{fig.MSE_evolve} that AMP-ABE and AMP-Wiener with fewer iterations already achieve NMSEs that are close to the NMSE shown in Table~\ref{tb.resize}.
In Figure~\ref{fig.MSE_evolve}, the horizontal axis represents iteration numbers, and the vertical axis represents NMSE.
It is shown in Figure~\ref{fig.MSE_evolve} that the NMSE drops markedly from $-10$ dB to $-21$ dB  for AMP-Wiener (solid line) and from
$-5$ dB to $-19$ dB for AMP-ABE (dash-dot line), respectively.
Note that the average NMSE is approximately $-21$ dB for AMP-Wiener and  $-19$ dB for AMP-ABE around iteration 15. Therefore, we may halve the runtimes of AMP-ABE and AMP-Wiener
(to approximately 1.7 seconds) by reducing the number of
AMP iterations from 30 to 15.

The simulation for the larger image patches extracted by Method 2 is slow, 
and thus the results for Turbo-BG and MCMC have not been obtained for 
Table~\ref{tb.192}. We believe that Turbo-BG is only slightly worse than 
Turbo-GM. At the same time, we did test for MCMC on several images, and 
found that the NMSEs obtained by MCMC were usually 
0.5 dB higher than AMP-Wiener and the runtimes of MCMC usually exceeded 
1,500 seconds. Similar to Figure~\ref{fig.MSE_evolve}, it can be seen from Figure~\ref{fig.MSE_evolveb} that the runtimes of our AMP based approaches could
be further reduced by reducing the number of AMP iterations without much deterioration in estimation quality.

\begin{table}[t]
\centering
\begin{tabular}{| c | c | c |}
\hline
Algorithm   & NMSE (dB) & Runtime (sec) \\
\hline 
  Turbo-GM~\cite{Som2012} & -19.64 & 56.12 \\

  AMP-ABE & -17.57 & 15.99 \\
    
AMP-Wiener & -20.29 & 15.53 \\
\hline
\end{tabular}      
\caption{\small\sl 
NMSE and runtime averaged over 591 images extracted by Method 2: a $192\times 192$ patch is extracted from the upper left corner of each image.
The number of measurements~$M=11,059$, 
and the measurements are noiseless.}
\label{tb.192}
\end{table}

\begin{figure}[t]
\vspace*{-0mm}
\begin{center}
\includegraphics[width=160mm]{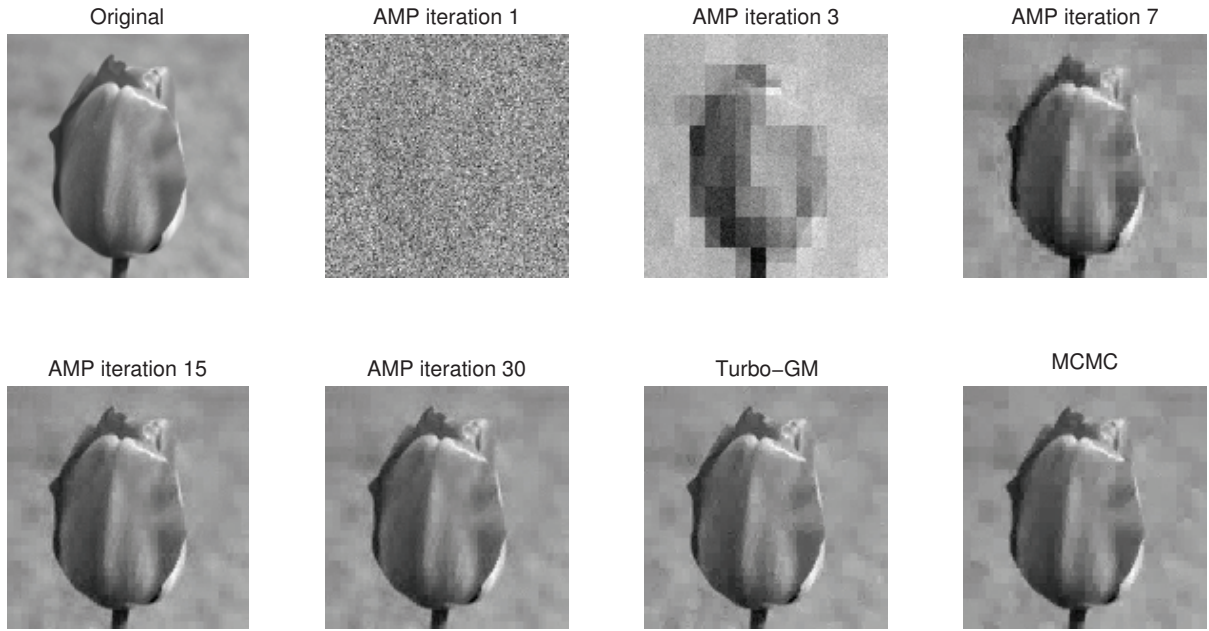}
\end{center}
\vspace*{-0mm}
\caption{\small\sl 
Original ``10\_5\_s.bmp" input image, the estimated images using AMP-Wiener at iterations 1, 3, 7, 15, 30, and the estimated images using Turbo-GM and MCMC. The image patch is extracted by Method 1: a $192\times 192$ patch is first extracted, and then resized to $128\times128$. 
The NMSE of each estimated image is as follows, AMP-Wiener iteration 1, $-11.46$ dB; AMP-Wiener iteration 3, $-18.17$ dB; AMP-Wiener iteration 7, $-26.28$ dB; AMP-Wiener iteration 15, $-30.07$ dB; AMP-Wiener iteration 30, $-30.36$ dB; Turbo-GM~\cite{Som2012}, $-29.62$ dB; MCMC~\cite{BCS2008spike}, $-29.13$ dB.}
\label{fig.Img_evolve}
\end{figure}

{\bf Result 2:}\
As a specific example, Figure~\ref{fig.Img_evolve} illustrates one of the 591 image patches
and the estimated patches using AMP-Wiener at iterations 1, 3, 7, 15, and 30. We also present the estimated patches using Turbo-GM and MCMC. It can be seen from Figure~\ref{fig.Img_evolve} that the estimated images using AMP-Wiener are gradually denoised as the number of iterations is increased, and the NMSE achieved by AMP-Wiener at iteration 15 already produces better reconstruction quality than Turbo-GM and MCMC.

\subsection{Performance of scalar denoisers}
\label{subsec.BM3D}

Having seen that AMP-Wiener consistently outperforms AMP-ABE, let us now understand why AMP-Wiener achieves lower NMSE than AMP-ABE. 

\begin{figure}[t]
\vspace*{-0mm}
\begin{center}
\includegraphics[width=78mm]{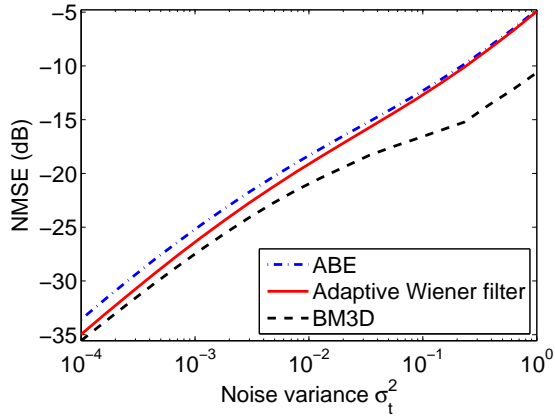}
\end{center}
\vspace*{-0mm}
\caption{\small\sl 
Average NMSE over 591 images versus noise variance. Image patches are extracted by Method 2: a $192\times 192$ patch is extracted from the upper left corner of each image. These image denoisers are applied to scalar channels~\eqref{eq.scalar}.}
\label{fig.varyNoise}
\end{figure}

We test for ABE~\cite{Figueiredo2001} and the adaptive Wiener
filter~\cite{Kivanc1999} as scalar image denoisers in scalar channels
as defined in~\eqref{eq.scalar}. In this simulation, we use the 591 image patches extracted by Method 2, and add i.i.d. Gaussian noise~$\mathcal{N}(0,\sigma^2)$ to the image patches. 
The pixel values are normalized to be between $0$ and $1$, and we verified from the simulations for Table~\ref{tb.192} that the estimated noise variances
of the scalar channels in AMP iterations are typically between
$1\times10^{-4}$ and $1$.
In Figure~\ref{fig.varyNoise}, the vertical axis represents NMSE, and the horizontal axis represents different noise variances varying from~$1\times10^{-4}$ to $1$ . It is shown in Figure~\ref{fig.varyNoise} that the adaptive Wiener filter (solid line) consistently achieves lower NMSE than
ABE (dash-dot line) for all noise variances, which suggests that 
AMP-Wiener outperforms AMP-ABE in every AMP iteration, and thus outperforms
AMP-ABE when we stop iterating in iteration 30. Therefore, in order to achieve favorable reconstruction quality, it is important to select a good image denoiser within AMP. 
With this in mind, we include the NMSE of the image denoiser 
``block-matching and 3-D filtering" BM3D~\cite{Dabov2007} in 
Figure~\ref{fig.varyNoise}, and find that BM3D (dashed line) has lower NMSE than the adaptive Wiener filter, especially when the noise variance is large. Note that the NMSEs of ABE for difference noise variances are 
within 1 dB of the NMSEs of the adaptive Wiener filter, but this 
performance gap in scalar denoisers is amplified to more than
2 dB (refer to Table~\ref{tb.192}) when applying the scalar denoisers within AMP. In other words, it is possible that applying BM3D within AMP
could achieve better reconstruction quality than AMP-Wiener. 
However, one challenge of applying BM3D within AMP will be that it is not clear whether the Onsager reaction term in~\eqref{eq.AMPiter2} can be computed in closed form or numerically, and thus an alternative way of approximating the Onsager reaction term may need to be developed.
During the review process of our paper, Metzler et al.~\cite{Metzler2014} showed how to compute the Onsager correction term numerically, thus allowing to use different image denoisers within AMP.

\subsection{Reconstruction quality versus measurement rate}
\label{subsec.varyM}

Finally, we also evaluate the performance of each algorithm by plotting the NMSE (average NMSE over 591 images) versus the measurement rate $R=M/N$. The measurement matrix ${\bf A}$ is generated the same way as the numerical setting in Section~\ref{subsec.runtime}.

\begin{figure}[t]
\vspace*{-0mm}
\begin{center}
\includegraphics[width=80mm]{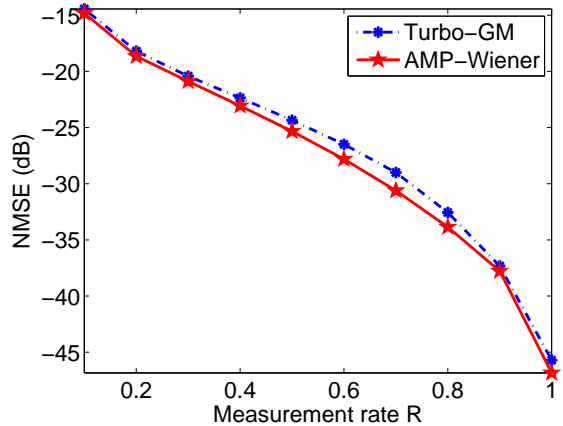}
\end{center}
\vspace*{-0mm}
\caption{\small\sl 
Average NMSE over 591 images versus measurement rate. Image patches are extracted by Method 1:
a $192\times192$ patch from the upper left corner of each image is first extracted, and then resized to $128\times128$.}
\label{fig.varyM_resize}
\end{figure}

\begin{figure}[t]
\vspace*{-0mm}
\begin{center}
\includegraphics[width=80mm]{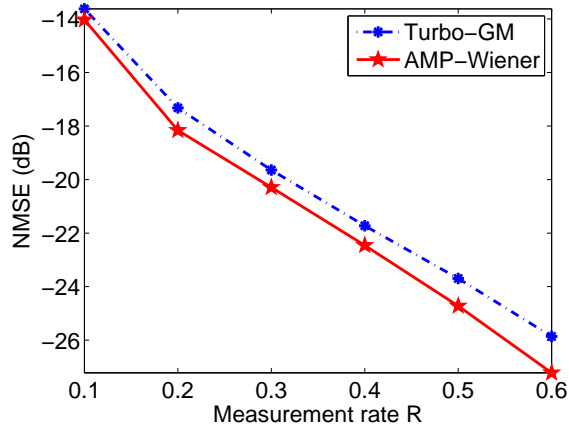}
\end{center}
\vspace*{-0mm}
\caption{\small\sl 
Average NMSE over 591 images versus measurement rate. Image patches are extracted by Method 2: a $192\times 192$ patch is extracted from the upper left corner of each image.}
\label{fig.varyM192}
\end{figure}

{\bf Result:}\
Figures~\ref{fig.varyM_resize} and~\ref{fig.varyM192} illustrate how the NMSEs achieved by AMP-Wiener and Turbo-GM vary when the measurement rate $R$ changes, 
where the horizontal axis represents the measurement rate $R=M/N$, and the vertical axis represents NMSE.
Figures~\ref{fig.varyM_resize} shows the results for image patches extracted by Method 1, and the measurement rate $R$ varies from $0.1$ to $1$. Figure~\ref{fig.varyM192} shows the results for image patches extracted by Method 2. Because the simulation for $192\times192$ image patches is relatively slow, we only show results for $R$ that varies from $0.1$ to $0.6$.
It can be seen from Figures~\ref{fig.varyM_resize} and~\ref{fig.varyM192} that AMP-Wiener (solid line with pentagram markers) achieves lower NMSE than that
of Turbo-GM (dash-dot line with asterisks) for all values of $R$.

\section{Discussion}
\label{sec:disc}

In this paper, we proposed compressive imaging algorithms that apply image
denoisers within AMP. Specifically, we used the ``amplitude-scale-invariant Bayes estimator" (ABE)~\cite{Figueiredo2001} and an adaptive Wiener 
filter~\cite{Kivanc1999} within AMP. Numerical results showed that AMP-Wiener achieves the lowest reconstruction error among all competing algorithms in all simulation settings, while AMP-ABE also offers competitive performance.
Moreover, the runtimes of AMP-ABE and AMP-Wiener are significantly lower than those of MCMC~\cite{BCS2008spike} and the Turbo approaches~\cite{Som2012}, 
and Figures~\ref{fig.MSE_evolve} and~\ref{fig.MSE_evolveb} suggest that the 
runtimes of our AMP based algorithms could be reduced further if we accept a slight deterioration in NMSE.

Recall that the input of the denoising function~$\eta_t$ in~\eqref{eq.AMPiter1} is a noisy image with i.i.d. Gaussian noise, and so we believe that any image denoiser that deals with i.i.d. Gaussian noise can be applied within AMP. At the same time, in order to develop fast AMP based compressive imaging algorithms, the image denoisers that are applied within AMP should be fast.
By comparing the denoising quality of ABE~\cite{Figueiredo2001} and the adaptive Wiener filter~\cite{Kivanc1999} as image denoisers in scalar channels, we have seen that AMP with a better denoiser produces better reconstruction quality for compressive imaging problems. With this in mind,
employing more advanced image denoisers 
within AMP may produce promising results for compressive imaging problems.
The development of such compressive imaging algorithms is left for future work.

This paper used squared error ($\ell_2$-norm error) as the performance metric, and AMP-Wiener was shown to outperform the prior art. Besides the $\ell_2$-norm error, Bayati and Montanari~\cite{Bayati2011} have shown that the SE in AMP also holds for other error metrics such as~$\ell_p$-norm error where $p\neq2$. We believe that, when some error metric besides $\ell_2$ error is considered, there exist denoisers that are optimal for this error metric of interest~\cite{Tan2014}, and thus by applying these denoisers within AMP we may be able to achieve optimal reconstruction results for matrix channels. The development of such denoisers is left for future work.

\section*{Acknowledgments}
We thank Phil Schniter for providing information about the numerical settings evaluated in his work with Som~\cite{Som2012}; Liyi Dai, Nikhil Krishnan, and Junan Zhu for useful discussions;  and the reviewers for their careful evaluation of the manuscript.

\ifCLASSOPTIONcaptionsoff
\newpage
\fi
\bibliographystyle{IEEEtran}
\bibliography{IEEEabrv,cites}

\end{document}